\documentclass[10pt, conference, compsocconf]{IEEEtran}
\ifCLASSINFOpdf
  \usepackage[pdftex]{graphicx}
\else
  \usepackage[dvips]{graphicx}
\fi
\usepackage[cmex10]{amsmath}
\begin{document}
%
\title{Personalized Expertise Search at LinkedIn}



%
\author{\IEEEauthorblockN{Viet Ha-Thuc \IEEEauthorrefmark{1},
Ganesh Venkataraman\IEEEauthorrefmark{1},
Mario Rodriguez\IEEEauthorrefmark{1}, 
Shakti Sinha\IEEEauthorrefmark{1},
Senthil Sundaram\IEEEauthorrefmark{1} and
Lin Guo\IEEEauthorrefmark{1}}
\IEEEauthorblockA{\IEEEauthorrefmark{1}Email:vhathuc,ghvenkat,mrodriguez,ssinha,ssundaram and lguo@linkedin.com\\
LinkedIn, 2029 Steirlin Ct, Mountain View, CA, USA}
}


\maketitle

\begin{abstract}
LinkedIn is the largest professional network with more than 350 million members. 
As the member base increases, searching for experts becomes more and more challenging. 
In this paper, we propose an approach to address the problem of personalized expertise search on LinkedIn, particularly for exploratory search queries containing {\it skills}. 
In the offline phase, we introduce a collaborative filtering approach based on matrix factorization. Our approach estimates expertise scores for both the skills that members list 
on their profiles as well as the skills they are likely to have but do not explicitly list. 
In the online phase (at query time) we use expertise scores on these skills as a feature in combination with other features to rank the results. 
To learn the personalized ranking function, we propose a heuristic to extract training data from search logs while handling position and sample selection biases.
We tested our models on two products - LinkedIn homepage and LinkedIn recruiter.
A/B tests showed significant
improvements in click through rates - {\bf 31\%} for CTR@1 for recruiter ({\bf 18\%} for homepage) as well as downstream messages sent from search -
    {\bf 37\%} for recruiter ({\bf 20\%} for homepage). As of writing this paper, these models serve nearly {\bf all live traffic} for skills search on
LinkedIn homepage as well as LinkedIn recruiter.

\end{abstract}


%
\IEEEpeerreviewmaketitle

\section{Introduction}

\begin{figure}
\centering
\includegraphics[width=0.45\textwidth]{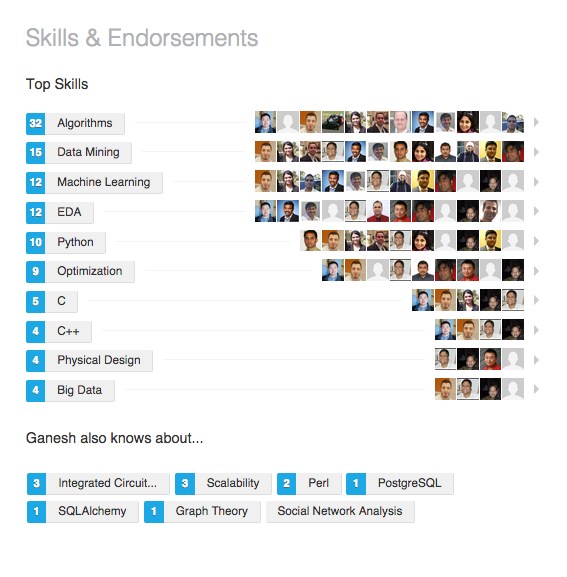}
\caption{Example of Skill Section with Endorsements}
\label{skills}
\end{figure}

The LinkedIn platform serves the professional networking needs of 350MM+
members worldwide. Members visit the site for various reasons including recruiting candidates, finding jobs, connecting with other people, reading professional content, etc. About {\it 62\%} of the company revenue is from Talent Solutions\footnote{http://blog.linkedin.com/2015/04/30/linkedins-q1-2015-earnings/}, which is a product helping recruiters and corporations around the world find the right talent. Thus, the problem of finding candidates with expertise in some certain areas is extremely important to LinkedIn. 

LinkedIn allows members to add  {\it skills} to their profile. Typical example of skills for a software engineer would be - ``Algorithms'', ``Data Mining'', ``Python'', etc. On LinkedIn, there are about 40 thousand standardized skills. Members can also {\it endorse} skills of other members in their network. 
Thus, skills are an integral part of members' profiles to help them showcase their professional expertise (see Figure \ref{skills}). 
In this paper, we focus on the problem of personalized expert ranking for queries containing one or several skills.

The area of expert finding was initiated in the field of knowledge management~\cite{Davenport1998}. 
It became a very active research area since the 
appearance of the Enterprise Track in Text Retrieval Conference (TREC). 
Over the last decade, many research papers have been published in this area. However, the task of expertise search on LinkedIn has unique challenges. 
\begin{enumerate}
\item Scale: With a member base more than 350 million and about 40 thousand skills, it is by far bigger than the benchmark data sets in the literature. 
For instance, TREC data set contains about 1100 candidates and 50 topics. 
\item Sparsity: Members may not list all the skills they have. The system should estimate expertise scores for the skills that members have 
    but do not explicitly list on their profiles. For example, if a member had
    listed ``hadoop'' and ``java'' explicitly, one could infer expertise scores
    for ``mapreduce'' as well.
\item Personalization: Expertise search on LinkedIn has a strong recruiting intent. Searchers not only care about candidates' expertise but also many personalized aspects.
These personalized aspects may include the social connection distance from the candidate, geographic location, etc.
\end{enumerate}

This paper describes our approach to addressing above mentioned challenges.
We propose a way to estimate expertise scores offline {\it at scale}. 
Our techniques, which involve collaborative filtering and matrix factorization {\it infer} expertise scores for skills that members have, including the ones that they do not even explicitly list in their profiles. 
When a searcher enters a query containing one or several skills, our ranking system
uses candidates' expertise scores on these skills as a feature in combination with other features. Our feature set contains personalized (location, social connection distance) as 
well as non-personalized (text matching) features. 
The ranking function is learned by leveraging learning-to-rank techniques. 
We detail practical issues faced when deploying learn-to-rank models in
personalized ranking at LinkedIn scale.
In particular, we detail our attempts to address position and sample selection biases while extracting training data for 
learning {\it personalized ranking} functions from search logs.

The techniques detailed in this work are used to rank a significant percentage of queries within LinkedIn. 
A/B tests were done on the LinkedIn homepage as well as LinkedIn recruiter search, a paid product.
On the homepage search, our proposed approach 
improves CTR@1 by {\it18\%},  CTR@10 by {\it 11\%} and number of emails sent from search page by {\it 20\%}.
On the premium recruiter search, our proposed approach improves CTR@1, CTR@10
and number of messages sent from search page by {\it 31\%}, {\it 18\%} and {\it 37\%}, respectively.
Note that while our examples are shown for skills related to software engineering/data mining, our approach is generic for 
any skills listed or inferred from our members profile. Our results are
validated across different domains including (but not limited to) sales,
marketing, engineering management etc. Currently, these models serve nearly {\bf all live traffic} for skills search on LinkedIn homepage as well as LinkedIn recruiter.

The rest of the paper is organized as follows. 
Section 2 reviews background on expert finding and learning to rank techniques for search ranking. 
Section 3 presents how we estimate skill expertise scores. Section 4 details how we learn a search ranking function by using the 
expertise scores in combination with other personalized and non-personalized features. 
We discuss experimental results in Section 5. Finally, concluding remarks can be found in Section 6.

\section{Background} \label{background}
\subsection{Expertise scores}
Expertise scores yield an ordered list of members for every skill. In other
words, it maps the tuple (member, skill) into a score. The closest match to published literature is in the
area of expert finding in the field of knowledge management \cite{Davenport1998}. 
This became a very active research area since the appearance of TREC 2005 and 2006 \cite{Craswell2005}, \cite{Soboroff2006}. 
In the track, the task of finding experts is defined as follows: given a collection of crawled documents including email, homepages, etc., a list of candidates and a list of topics, 
the task is to rank the candidates given each topic. Almost all of the systems submitted to TREC Enterprise Tracks in 2005 and 2006 
and subsequent research (\cite{Balog2006},~\cite{Balog2012},~\cite{Fang2007},~\cite{Petkova2008}) view this task as a text retrieval problem. 
They fall into one of the two approaches: candidate-based approach which ranks candidates by textual similarity between the topics and candidates' profiles and document-based approach which first retrieves documents relevant to the topics, 
then discover candidates mentioned in the documents and estimate their associated scores. 
Besides these approaches, there is also research on combining multiple evidence associating candidates to topics 
~\cite{Macdonald2008}~\cite{Yang2009}~\cite{Hofmann2010}. Focusing on expert finding on social graphs, Zhang et al.~\cite{Zhang2007}, Rode et al.~\cite{Rode2007} 
propose a two-step process: (i) using language models or heuristic approaches to compute an initial expertise score for 
each candidate, and (ii) using graph-based algorithms to propagate scores computed in the first step and re-rank experts.

Compared to the previous research, our work on skill expertise scores is different in the following aspects. 
First, our system leverages the unique data source of LinkedIn including member profiles containing different sections, such as work experience, 
education, skills, projects, social endorsements and the high-level signals derived from the profile, such as seniority and popularity. 
The second unique aspect is scale. Our system estimates expertise of more than 350 million members on about 40 thousand skills, 
which is by far bigger than the benchmark datasets provided by TREC and others. For instance, 
TREC dataset contains about 1100 candidates and 50 topics. Finally, previous research typically ranks experts for 
one topic at a time and does not take into account topic co-occurrence  patterns. Our collaborative filtering technique is able to infer expertise scores for the skills that members do not even list in their profiles based the skills they do and 
relationships amongst skills.

\subsection{Learning to Rank}
The second phase in our work that combines skill expertise scores with other signals into a final search ranking 
is related to the area of learning to rank for search. Learning to rank has been a key problem for information retrieval systems and 
Web search engines in particular since these systems typically use many features for ranking and that makes it very 
challenging to manually tune the ranking functions. 
There has been a lot of research published in the area of learning to rank, typically falling into one of the 
three following categories. Pointwise approaches view ranking as traditional binary classification or regression problems. 
Pairwise approaches take input as a set of pairs of documents in the form of one document is more relevant than the other with respect to a specific query. 
These approaches then learn a ranking functions minimizing the number of incorrectly ordered pairs. The state-of-the-art for learning to 
rank is composed of listwise approaches. These approaches typically view the entire ranked  list of documents as a learning instance 
while optimizing some objective function defined over 
all of the documents, such as mean average precision (MAP) or normalized discounted accumulative gain (NDCG) \cite{Cao2007}. 
We refer the readers who are interested in more details of learning to rank to~\cite{Chapelle2011}~\cite{Chapelle2011b}~\cite{Liu2010}~\cite{Li2011} for more comprehensive reviews. 
 
A key element of learning to rank is collecting ground truth data. Traditionally, ground truth data is labeled by 
either professional editors or crowd sourcing~\cite{Chapelle2011}. 
Given a pair of query and document, human judges give either a binary judgment (relevant or not-relevant) or graded categories, such as perfect, 
excellent, good, fair or bad. 
However, the issues with this approach are (i) It is expensive and not scalable and 
(ii) It is very hard for the judges to judge the relevance on behalf of some other user, 
making it challenging to apply the approach for personalized ranking. 
For these reasons, some previous research proposes to extract labeled data using click logs~\cite{Joachims2002}~\cite{Carterette2007}~\cite{Craswell2008}~\cite{Hofmann2014}. 
Nonetheless, collecting training data from click logs presents its own challenges. 
We will have to find a way of handling position bias and sample selection bias. User eyes tend to scan from top results to bottom and some of the bottom results may not even be viewed. 
Thus, marking unclicked results that appear low in the original ranking may result in unfair penalization. 
Moreover, user clicks generally occur on the top ranked results and it quickly tapers down. 
So, the documents in the training data are a very biased sample of all results. 
Radlinski and Joachims \cite{Radlinski2006} propose an approach called FairPair to collect labeled data from click logs and avoid the position bias. 
Nevertheless, the labeled data is mainly for pairwise learning to rank algorithms, not particularly suitable with the listwise ones. Second, it still suffers the sample selection bias.

\section{Skill Expertise Scores}
The skill expertise algorithm infers skill expertise scores, which can be thought of as the probability that a member is an expert given a skill query containing one of more standardized skills, or $p(expert | \mathbf{q}, member)$.  We discuss what we mean by \emph{expert} in section \ref{preliminary_learner}.  

Our approach enables us to score members for multi-skill queries, $\mathbf{q} = (\text{``java''}, \text{``data mining''}, \text{``information retrieval''})$, as well as single-skill queries, $q = (\text{``java''})$.  This enables us to rank members given the context of a skill-only query for all members $M$, where $|M| = m$, given any arbitrary skill combination from the standardized skill set $S$, where $|S| = s$.  We distinguish multi-skill queries from single-skill queries in our notation by bold facing the query vector.

\begin{figure*}
\centering
\includegraphics[width=0.7\textwidth]{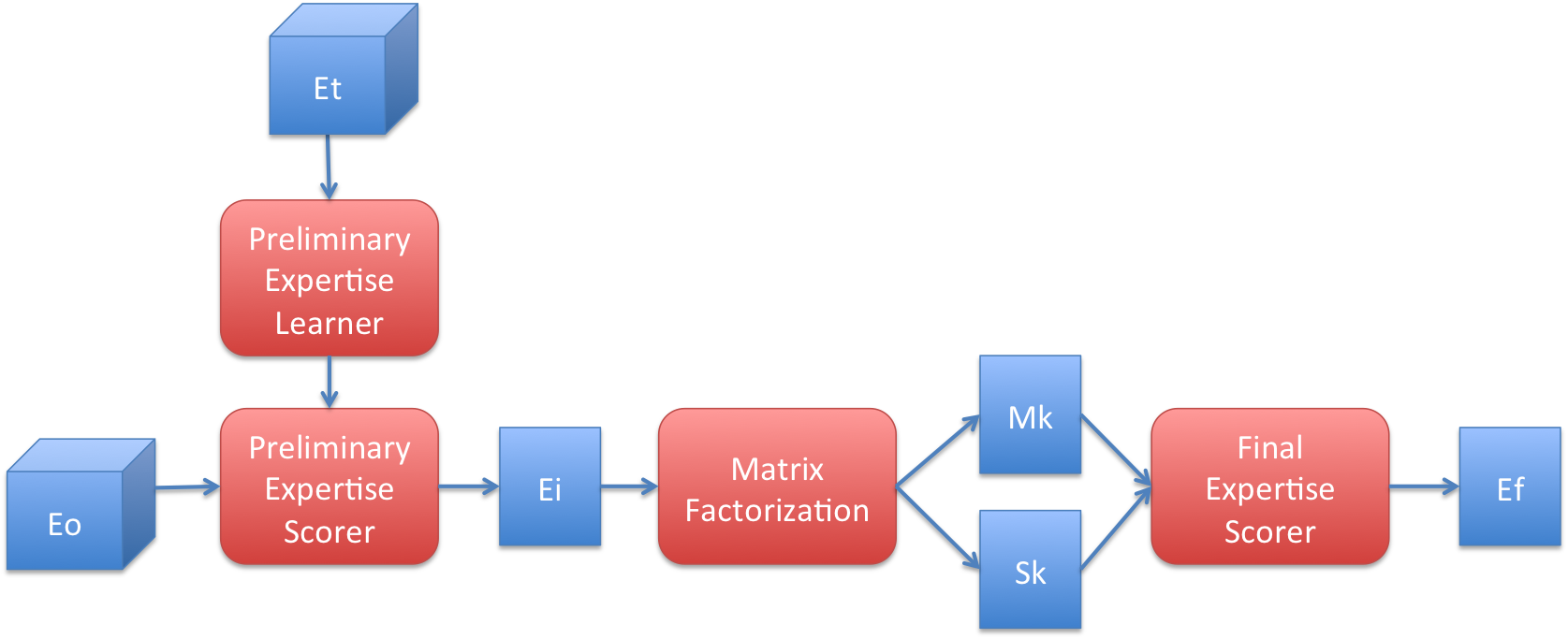}
\caption{Skill expertise work flow.  $E_o$, of which $E_t$ is a subset, represents a 3-rd order tensor of dimensions $m \times s \times f$ which denotes an f-dimensional feature vector for every member-skill pair.  The \emph{Preliminary Expertise Scorer} converts the f-dimensional feature vector into an expertise score, yielding a sparse skill expertise score matrix $E_i$ of dimensions $m \times s$.  The \emph{Matrix Factorization} component factorizes $E_i$ into dense factor matrices $M_k$ and $S_k$, which when multiplied together yield matrix $E_f$, a denser and final expertise matrix.}
\label{sys_diag}
\end{figure*}

To compute these skill expertise scores for arbitrary queries, we use a 2-step process illustrated in figure \ref{sys_diag}, which is composed of: (1) a supervised learning step to compute $p(expert | q, member)$ which outputs an initial, sparse, expertise matrix $E_i$, of dimensions $m \times s$, and (2) a matrix factorization step that factorizes $E_i$ into latent factor matrices, $M_k$ and $S_k$ of dimensions $m \times k$ and $s \times k$ respectively, which when multiplied, allow us to compute a final, denser, expertise matrix $E_f$.  Here, $k$ represents the size of the latent reduced-dimensionality space onto which both, members and skills, are projected, and where $k \ll m$ and $k \ll s$.  

Matrix $E_i$ is the interface between the 2 steps in the pipeline.  Having this interface explicitly formalized allows for development on both pipeline stages and algorithms to be independent.  We now explore the most salient pipeline components in more detail.

\subsection{Preliminary Expertise Learner and Scorer} \label{preliminary_learner}
Expertise is the umbrella term we use to collectively refer to various attributes of members that capture relevant characteristics of those expected to rank high for a given skill query. These attributes are measured using various signals available on the site, many of which rely on complex algorithms for standardization and inference:

\begin{itemize}
  \item Seniority: we have many ways of measuring seniority, for example, as years of experience from a common reference point (graduation), or as years of experience typically required to reach someones current job title.
  \item Popularity: which may include measures of number of page views received, or number of endorsements received, as well as more elaborate derived signals such as what can be obtained from running PageRank \cite{ilprints422} on the page view graph or the endorsement graph.
  \item Influence: members have the ability to publish long-form content on the site, and this content is consumed by other members who may choose to further share it with their connections.  Members whose content generates a lot of engagement tend to be considered influential.
  \item Authority: members typically belong to organizations that are hierarchical in nature, and this hierarchy is often reflect on job titles by means of qualifiers, such as \emph{associate}, \emph{senior}, \emph{principal}, \emph{manager}, \emph{director}, or \emph{C-level managers} (e.g. CEO, CFO, CTO).
  \item Desirability: members are often contacted about job opportunities that may be of interest to them, and some members may be contacted more often than others.  This variability in demand from hiring managers or recruiters depicts a measure of demand for a particular set of skills, educational background, or experience.
  \item Relevance: we also take into account the relevance of a given standardized skill to a member's profile.  This signal is derived from the same algorithm that generates recommendations for members to add skills to their profile (see \cite{Bastian2014}).  Note that we differentiate between how relevant a given skill is to a profile, vs. the expertise of the profile's owner at the given skill.  For example, the skill \emph{java} is likely relevant to the profile of a recent college graduate out of a Computer Science program, just as it is relevant to the profile of James Gosling, the father of the Java programming language, but both have different levels of expertise in Java.
\end{itemize}


These attributes become features associated with any given (member, skill) pair, where these pairs are generated from those skills that members list explicitly on their profiles which are above a certain relevance threshold.  The feature vectors associated with each (member, skill) pair are consolidated in a 3-rd order tensor of dimensions $m \times s \times f$, which is depicted as $E_o$ in Figure \ref{sys_diag}, where $F$ is the set of all possible features associated with a (member, skill) pair and $|F|=f$.  A subset of $E_o$ is used as training data, denoted $E_t$ and is discussed in the next section.

\subsubsection{Training Data} \label{skill_train_data}
Our Preliminary Expertise Learner is based on a classifier that learns to distinguish between our positive and negative examples, or more specifically, expert (member, skill) pairs vs. non-expert (member, skill) pairs.

The positive examples are sourced from various lists of experts available online: conference speakers (e.g. Strata), open-source committers (e.g. Apache Software Foundation committers), influential author lists (e.g. LinkedIn influencers), as well as cohorts of members about whom we have an expectation of their expertise (e.g. very senior members, in-demand members, etc).  Once we have identified these expert members, and are able to link them with a profile, we extract their explicitly listed skills and use various heuristics to filter their list of explicit skills into a smaller list that may be considered to be their most relevant skills.  An example of such heuristic might be to filter out skills with a profile relevance score less than some threshold $t$.  This is done to filter out potential outlier skills added in jest.

The negative examples are similarly generated based on various strategies.  Examples of non-expert (member, skill) pairs may include regular members paired with random skills, or skills that are only mildly relevant to them, as well as members with profiles tagged as \emph{spam} or \emph{fake} (with a certain probability), and skills associated with them.

Finally, the positive and negative examples of (member, skill) pairs are validated using crowd sourcing. The final set is then divided into three subsets: training, testing, and validation sets.  Training and testing is discussed in the next section, and validation is discussed in section \ref{validate_expert_scores}.

\subsubsection{Learning}
We employ a logistic regression model for learning to distinguish between the expert and non-expert (member, skill) pairs using the training data described in the previous section, the primary reason for using a simple linear model being explainability.

The test set is used to evaluate the accuracy of the logistic regression model, as well as to calibrate meta-parameters, such as regularization parameters and cohort mix-proportion in the training set.

\subsubsection{Scoring}
At this point in the pipeline, we use the logistic regression model to estimate $p(e | q)$ for those standard skills which members have listed explicitly on their profiles, with the further consideration that those skills need to be highly relevant to the profile in question (so, Karaoke singing for a Director of Search Quality would likely not make the cut).  

This gives us a very sparse matrix, denoted as $E_i$ in figure \ref{sys_diag}, where most of the values in the $m \times s$ matrix are considered to be unknown.  In fact, since at the time of this writing, we limit users to listing up to 50 explicit skills on their profiles, and given that we have about $\sim40,000$ standardized skills, even if all $\sim$ 350MM members list the maximum number of skills (assuming they are all standard skills), we have a matrix with 99.9\% unknown values.

\subsection{Matrix Factorization}
In the second stage of the pipeline, we aim to \emph{discover} some of those unknown values in $E_i$ by leveraging collaborative filtering techniques that enable us to uncover insights such as \emph{users who are good at Statistics also tend to be good at Data Mining}. 

\subsubsection{Normalization}
In order to decouple the factorization algorithm from the particulars of the supervised learning step, prior to factorization, we apply a rank-based inverse normal transformation (see \cite{rankit}) to those scores in $E_i$ which are known (as well as shift and scale the standard normal scores to be non-negative with an arbitrary mean and standard deviation).

\subsubsection{Factorization}
When factorizing the normalized $E_i$ matrix, we sought to find the factors that optimized a loss function similar to equation 3 in \cite{implicit_cf}, with a slight modification:

\begin{equation}
\min_{x_{*},y_{*}} \sum\limits_{m,s} c_{ms}(s_{ms} - x_{m}^T y_{s})^2 + \lambda( \sum\limits_{m} ||x_m||^2 + \sum\limits_{s} ||y_s||^2)
\label{eq:fact_loss}
\end{equation}

Where the goal is to find a vector $x_m \in \rm I\!R^k$ for each member $m$, and a vector $y_s \in \rm I\!R^k$ for each skill $s$ that will factor member-skill affinities.  In other words, skill expertise scores are assumed to be the inner products: $s_{ms} = x_{m}^T y_{s}$.  We refer to these vectors as the member-factors and skill-factors, respectively.  This formulation is different from \cite{implicit_cf} in that we use the actual normalized expertise score in $s_{ms}$, rather than simply using an indicator variable denoting whether or not the member has the skill.

Note that in this formulation, similar to \cite{implicit_cf}, we account for \emph{all} (member, skill) pairs, $(m,s)$, rather than only those which are known.  We treat the unknown scores as zeros with low confidence, and the not unknown scores as being high-confidence values.  The confidence function is similar to the one used in \cite{implicit_cf}:

\begin{equation}
    c_{m,s}= 
\begin{cases}
    \alpha,& \text{if } s_{ms} > 0\\
    1,              & \text{otherwise}
\end{cases}
\label{eq:fact_confidence}
\end{equation}

The exact values for $\lambda$ (the regularization parameter), $k$ (the size of the latent reduced-dimensionality space), and $\alpha$ (the high-confidence value for known expertise scores) were determined by cross-validation on the reconstructed matrix.

The reason for this loss function is we know that most members typically only specialize in a very small subset of the entire 40,000-dimensional skill vector space, so \emph{most} of their scores should in fact be zero.  This formulation is similar to weighted least squares regression (see \cite{wlsr}), where less weight is given to less precise measurements and more weight is given to more precise measurements when estimating the unknown parameters of the model.

\smallskip

\subsection{Final Expertise Scorer}
Having the latent representation of members and skills, we are free to reconstruct the entire expertise matrix.  However, we only do so for (member, skill) pairs with high relevance scores between the member's profile and the skill as described in Section \ref{preliminary_learner}

To compute the ranking for a single-skill query, we only need the expertise scores for that skill for members of interest, which is computed by using the inner product of the latent factor vectors $s_{ms} = x_{m}^T y_{s}$.   To compute the ranking for a multi-skill query, $\mathbf{q} = (s_1, s_2, s_3)$, we can do the same once we have the latent factor representation of the multi-skill query.  We define the latent factor representation of the multi-skill query to be a linear combination of the latent factors of each of the skills in the query, $y_{\mathbf{q}} = y_{s_1} + y_{s_2} + y_{s_3}$.  This enables us to compute the expertise score for the multi-skill query for a given member $m$, or $s_{m\mathbf{q}} = x_{m}^T y_{\mathbf{q}}$.

Rather than projecting the query to the latent factor space at query-time, we can take advantage of the distributive property of vector multiplication over addition so that at query-time, we need only sum the scores for those members satisfying the retrieval requirements of a given query:

\begin{equation}
s_{m\mathbf{q}} = x_{m}^T \cdot y_{\mathbf{q}} 
\label{eq:vec_dist_1}
\end{equation}

\begin{equation}
s_{m\mathbf{q}} = x_{m}^T \cdot (y_{s_1} + y_{s_2} + y_{s_3})
\label{eq:vec_dist_2}
\end{equation}

\begin{equation}
s_{m\mathbf{q}} = x_{m}^T \cdot y_{s_1} + x_{m}^T \cdot y_{s_2} + x_{m}^T \cdot y_{s_3}
\label{eq:vec_dist_3}
\end{equation}

To ensure fast retrieval and scoring, the skill expertise scores (e.g. $x_{m}^T \cdot y_{s_1} , x_{m}^T \cdot y_{s_2} , x_{m}^T \cdot y_{s_3}$) are precomputed offline and encoded as payloads in a Lucene~\cite{lia} inverted index, so that they are readily associated with each of a given member's skills.  In this setting, a member ID is a document ID in the inverted index, and the member's skills are terms in that document, and associated with each (doc ID, term ID) pair is a payload-encoded expertise score.

\subsection{Validation on Expertise Scores} \label{validate_expert_scores}

\begin{figure}
\centering
\includegraphics[width=0.45\textwidth]{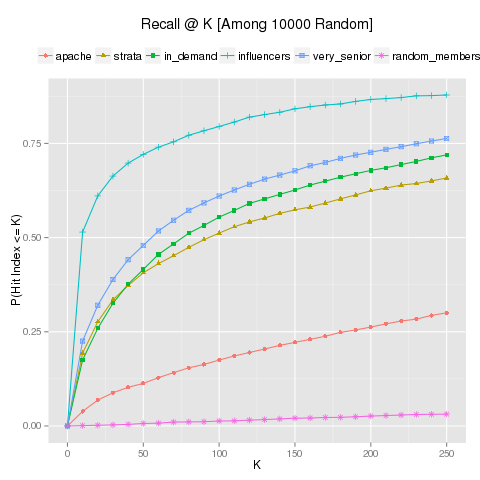}
\caption{The plot shows the probability that a member of the specified cohort (e.g. Strata speaker) is ranked at a position $i <= K$, as $K$ varies from 1 to 250.  We summarize the performance of the algorithm for a given cohort by using the \emph{area under the curve} (AUC) for a given K, as shown in table~[\ref{table_auc}].}
\label{expertise_algo_validation}
\end{figure}

For our validation step, we follow the same methodology discussed in~\cite{topn_rec_tasks}, 
using a hold-out validation dataset from our training dataset discussed in section~[\ref{skill_train_data}], see figure ~[\ref{expertise_algo_validation}] and table~[\ref{table_auc}].  This methodology allows us to use the AUC@K for each of the cohorts to compare different versions of the skill expertise model pipeline.

Each member cohort (e.g. influencer, strata, etc) represents a set of members about whom we have an expectation of how they should rank for a given skill query.  For example, given a seed member who is a Strata speaker and given the list of skills in which the seed member is considered an expert at (determined as described in section \ref{skill_train_data}), we use his/her top-skill(s) as a skill query for which we generate a ranking of members.  After we rank all members with regards to the query, we sample 10,000 relevant members, which, together with the seed member, becomes a ranked list of 10,001 for a specific skill query.  This methodology is repeated many times over for each of our member cohorts, for multi-skill and single-skill queries, to generate the plot in figure~[\ref{expertise_algo_validation}].  The plot shows the probability that the seed member is ranked at a given $K$, as $K$ varies from 1 to 250.  We summarize the performance of the algorithm for a given cohort by using the \emph{area under the curve} (AUC) for a given K, as in table~[\ref{table_auc}].

An advantage of this strategy is that it is completely independent of whatever algorithm was used to compute the skill expertise scores.

\begin{table}
\centering
\begin{tabular}{|l|c|} \hline
\bf{Cohort} & \bf{AUC@250} \\ \hline
Influencer & 0.76 \\ \hline
Very senior & 0.59 \\ \hline
In-demand & 0.53 \\ \hline
Strata & 0.50 \\ \hline
Apache & 0.19 \\ \hline
Random & 0.02 \\ \hline
\end{tabular}
\caption{AUC @ K=250, see figure~[\ref{expertise_algo_validation}].}
\label{table_auc}
\end{table}

\section{Learning Expertise Ranking Function}
In this section, we present how we leverage learning-to-rank algorithms to learn a ranking function for expertise search. The ranking function combines skill expertise scores described in the previous section as well as other features. 
Section 4.1 gives a description on the features in the ranking function. Section 4.2 details the learning algorithm as well as the proposed approach to extract training data from search logs and how to alleviate position and sample selection biases in the log data. 

\subsection{Features}
In this section, we give an overview of the features used in the ranking function for personalized expertise search. Due to business sensitivity, we cannot give details of how the features are computed. Instead, we give a high-level description. Most of the features are generally divided into the following categories.

\textbf{Expertise scores} 
As described in Section 3, we estimate an expertise score for every (member, skill) tuple.
When a searcher enter a query like ``java python senior developer", we first use a tagger to extract standardized skills mentioned in the query (``java" and ``python" in this example). on distributed computing platforms like Hadoop, allows processing very large datasets with complex methods e.g. matrix factorization to produce sophisticated signals like expertise scores. 
Detailed description of the tagger is beyond the scope of this paper.
Interested reader may refer to~\cite{Tan2008} for more information.
For every search result, we compute sum of his or her expertise scores on the skills.

\textbf{Textual features} 
The most traditional type of features in information retrieval are textual features (e.g. term frequency). These features are computed on different sections of user profiles, such as position titles, position descriptions, specialty, etc.  

\textbf{Geographic features (personalized features)} Expertise search on LinkedIn is highly personalized. For instance a simple
query like ``software developer" from a recruiter will produce very different
results when the searcher is in New York City as opposed to Sydney. Location plays an important role in personalizing the results. We
created multiple features capturing this.

\textbf{Social features (personalized features)} Another important aspect of personalization is to capture how the results socially relate to the searcher. 
We leverage a variety of the signals on LinkedIn, such as common friends, companies, groups and schools to generate the features in this category.

\subsection{Training Data}
Generally, the objective of learning to rank is to find a ranking function that computes a score representing the relevance between a document and query. 
In case of personalized ranking in social networks, the ranking function computes relevance score for every triple of (query, document, \textit{searcher}). As discussed in Section 2, a traditional way of generating labeled data by editors or crowd sourcing is not particularly suitable for personalized relevance at a large scale. Thus, we chose to collect training data using search logs. Process of collecting training data from search logs, however, presents its own set of challenges. We will have to handle two kinds of biases: position bias and sample selection bias.

\begin{figure}
\centering
\includegraphics[width=0.41\textwidth]{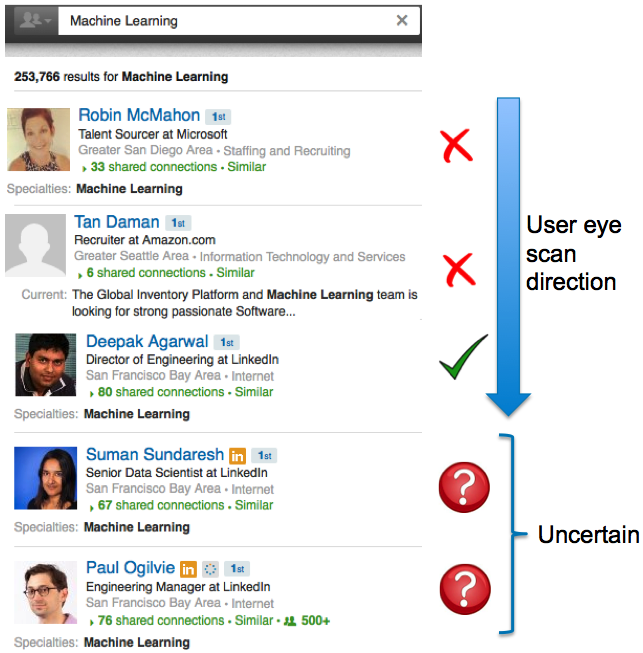}
\caption{Typically, a searcher's eyes scan from top results to bottom ones. For example, if the searcher skips results at positions 1 and 2 and clicks on result 3, then we assume that the clicked result is relevant and the skipped results are not.  We cannot be certain about the relevance judgement for results 4-5, since the searcher may not have looked at them.} 
\label{position_bias}
\end{figure}

Position bias happens when the likelihood of user actions e.g. clicks on search results is not only influenced by the relevance of the results but also their positions in the ranking~ \cite{Joachims2005}. A standard way to avoid the position bias is to randomize top-\textit{N} results, show them to a small percentage of traffic then collect feedback. 
However, this approach would change the rankings completely after each reload thereby leading to bad search experience. 
In this work, we use a pseudo-randomization technique: we first apply a raking function to rank all of the results, then re-rank the top-\textit{N} by a deterministic hash function on member Id, which is independent to result relevance. Thus, the results are stable and the top-\textit{N} rankings are still orthogonal to all of the features. Given the rankings, typically searcher eyes scan from top results to bottom ones. For example, if the searcher skips results at positions 1 and 2 and interacts with result 3 (see Figure \ref{position_bias}), then we assume that result 3 is relevant and the skipped results are not. Also note that results 4 and 5 are also shown to the searcher, but we cannot be certain if he or she looked at them and thinks they are not relevant or stops scanning at result 3. So, we simply ignore all of the results ranked bellow the last interacted one in the result page. We use \textit{graded relevance} for different searcher actions. For instance, if the searcher sends a message to, clicks or ignores a result, this result has a relevance label of two (perfect), one (good) or zero (bad), respectively.

The results on which we have user feedback are usually the top ones in the rankings. 
Thus, the documents in the training data is a biased sample of the overall document space. This can have adverse consequences in training. For instance, let us consider spam feature which indicates if a result is spam or not. Since this is a very important feature and the original ranking function gives a high weight for it, all of the top results are likely to be non-spam. So, when the ranking function is re-trained from the user feedback, the learner (mistakenly) learns that this feature has little discriminative power since all of the training instances have the same value. As a result, it gives a zero weight for this crucial feature. To address this issue, presumably the original ranking is not too bad in the sense that the tail results (very low ranked ones) are generally worse than the top ones, we randomly sample \textit{easy negatives} at the tails of the rankings. These results are still in the set of retrieved documents for the query but very unlikely to get clicked if they were presented. Thus, we consider these easy negatives as bad results and include them into the training set. Introducing easy negatives into the training data reduces the skewness of feature value distributions, particularly for dominant features  such as spam. Therefore, it has an effect of diversifying training pools and reducing sample selection bias.

We count every query issue as a unique search. For instance, if the query ``machine learning'' was issued five times, they are considered as five distinct searches. The reason for this is that even though the skill being searched for is identical across the five searches, they might have had different search contexts when considering other dimensions such as searcher or location etc. Moreover, this keeps the query distribution in the training data the same as in our live search traffic.

\subsection{Model Training}
Given a training dataset \textit{D}, we apply coordinate ascent algorithm, a popular optimization algorithm to search for a solution (feature weights) \textit{\textbf{w*}} that optimizes normalized discounted accumulative gain (NDCG) defined on the graded relevance labels as described above over \textit{D}. 


Without loss of generality, we assume all features are positively correlated with label (if a feature is negatively correlated with label, simply negate it). With this assumption, all of the feature weights are guaranteed to be positive, thus every solution \begin{math} \textbf{w} \in  {R^+}^N \end{math}, where \textit{N} is the number of features, can be mapped to an equivalent one (in terms of ranking) \begin{math} \boldsymbol{\lambda} \in P^N \end{math} (see Equation~[\ref{eq:weight_mapping}]), where \begin{math} P^N \end{math} is a multinomial manifold as described in Equation ~[\ref{eq:multinomial_manifold}]. The benefit of doing this is that all of the rank-equivalent solutions in \begin{math} {R^+}^N \end{math} is reduced to a single solution in the new space. Thus, the parameter space and as a result the number of local optima are significantly reduced.    

\begin{equation}
\lambda_i = \frac{w_i}{\sum_{i=1}^{N} w_i}\\
\label{eq:weight_mapping}
\end{equation}

\begin{equation}
P^N = \{\boldsymbol{\theta} \in R^N : \forall i  \: \theta_i \geq 0, \sum_{i=0}^{N} \theta_i=1\}
\label{eq:multinomial_manifold}
\end{equation}

The coordinate ascent algorithm optimizes the multivariate non-smooth objective function above by iteratively solving multiple one-dimensional optimization problems. At a time, it searches for optimal weight for a feature while keeping the others fixed and repeats the process. Randomization technique is used during the search process to avoid getting stuck in a local optimum. This algorithm has also shown to be efficient for learning search ranking functions in some other domains \cite{Metzler2007}.

After training different models (ranking functions) resulting from variety of parameter settings, we need to evaluate these models offline before deploying them in production. We use a held-out set from the training data as a test set. The models are used to re-rank the test data and evaluated by NDCG@K where \textit{K} equals to \textit{10} for LinkedIn homepage search and \textit{25} for recruiter search (these two products will be described in the next section). Those are the numbers of results shown in the first result page of the two products. We also compare coordinate ascent algorithm with other popular LTR algorithms in the literature on the test data. It turns out that coordinate ascent algorithm achieves the best  balance between performance and efficiency (scoring time) for our case. Since the paper does not focus on comparing different LTR algorithms (instead it emphasizes on practical challenges faced while deploying LTR in personalized ranking at an industrial scale), we do not show offline evaluation here to save space. Given offline performance, we take a few top models for online A/B tests in production. 

%

\section{Online Search Experimental Results}
\begin{figure}
\centering
\includegraphics[width=0.39\textwidth]{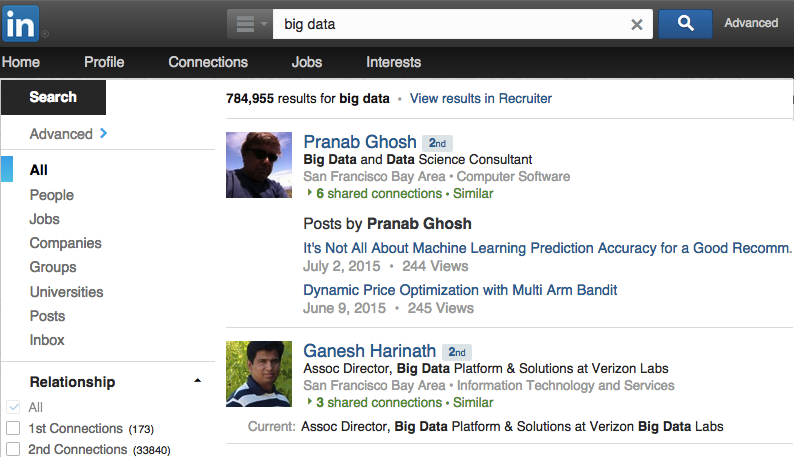}
\caption{LinkedIn Homepage Search}
\label{homepage_search}
\end{figure}

\begin{figure}
\centering
\includegraphics[width=0.39\textwidth]{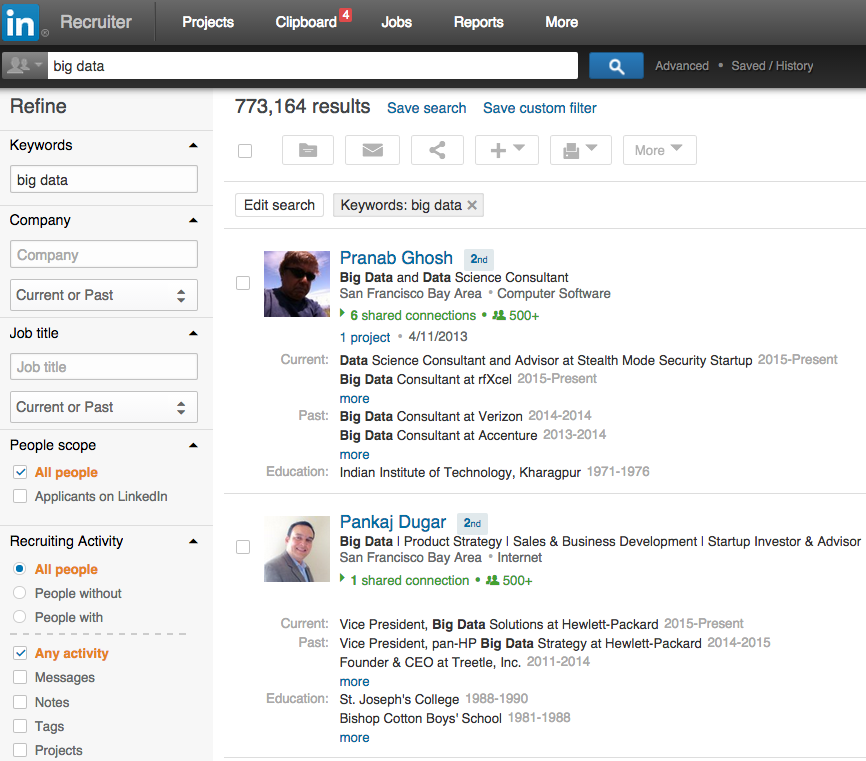}
\caption{LinkedIn Recruiter Search}
\label{recruiter_search}
\end{figure}

In this section, we verify effectiveness of the ranking functions by using A/B tests on live search traffic. 
We run the A/B tests on both of LinkedIn homepage search as well as LinkedIn recruiter search. 
The homepage search (See Figure~\ref{homepage_search}) offers every member the ability to discover people in their networks, 
subject to visibility rules. The recruiter search (See Figure \ref{recruiter_search}) is an enterprise product
where licensed recruiters use to search for candidates in the entire LinkedIn member base. 
Compared to recruiter search, the user base of homepage search is bigger and much more diverse. 
Recruiter search users are significantly more active.

\textbf{Baseline}: 
We compare the machine-learnt functions described previously with the legacy ranking functions. 
The legacy functions for both homepage search and recruiter search use all of the features described in Section 4.1 except expertise scores. 
Even though the weights of the features were manually tuned, the functions have been running on live traffic for a relatively 
long time and have been iteratively refined. The weights were tuned over
several years and several A/B tests.    

We run A/B test on both products for a reasonably long period of time (six weeks) to remove any novelty effect. 
When a searcher enters a query, we first use a tagger to tag query keywords with labels like {\it personal name (first and last name)}, 
{\it skill}, {\it location}, {\it job title}, etc. For detailed description of the tagger, we refer interested readers to~\cite{Tan2008}. 
Since we are interested in exploratory expertise search, we focus on queries containing at least one skill and not containing personal names. 
For searches satisfying this condition, the traffic is randomly split into treatment and control buckets. 
Searches in the treatment buckets are ranked by the machine-learnt functions. Control buckets use legacy ranking functions (our baseline). During six weeks, each bucket on either product contains several hundreds of thousand skill searches.

Based on members' interactions with search results, we compute two sets of metrics. 
The first one is so called first-order metrics including click-through-rate at the first results (CTR@1), 
CTR at top ten results (CTR@10) and mean reciprocal rank (MRR). 
On both products, our final goal is not simply to have searchers click on results and view their profiles, 
but reach out to those results and eventually hire them. 
To partially capture this, we use a downstream metric measuring the number messages sent per search. 
We compare the treatment and control buckets on the metrics. 

Table \ref{table_homepage_first_order} shows first-order metrics on the homepage search of the baseline and the machine-learnt model having the best offline performance. Compared to the baseline, 
the new ranking function improves {\it18\%}, {\it11\%} and {\it 14\%} in terms of CTR@1, CTR@10 and MRR, respectively. 
Similarly, Table \ref{table_recruiter_first_order} shows that on LinkedIn recruiter search, the machine-learnt ranking function is {\it31\%}, {\it18\%} and {\it 22\%} 
better than the baseline in terms of CTR@1, CTR@10 and MRR.

\begin{table}
\centering
\begin{tabular}{|l|c|} \hline
& Improvement of Treatment over Control \\ \hline
CTR@1 & +18\% \\ \hline
CTR@10 & +11\% \\ \hline
MRR & +14\% \\ \hline
\end{tabular}
\caption{First order metrics on LinkedIn homepage search.}
\label{table_homepage_first_order}
\end{table}

\begin{table}
\centering
\begin{tabular}{|l|c|} \hline
& Improvement of Treatment over Control \\ \hline
CTR@1 & +31\% \\ \hline
CTR@10 & +18\% \\ \hline
MRR & +22\% \\ \hline
\end{tabular}
\caption{First order metrics on LinkedIn recruiter search.}
\label{table_recruiter_first_order}
\end{table}

\begin{table}
\centering
\begin{tabular}{|l|c|} \hline
& Improvement of Treatment over Control \\ \hline
Homepage Search & +20\% \\ \hline
Recruiter Search & +37\% \\ \hline
\end{tabular}
\caption{The number of downstream messages per search.}
\label{table_second_order}
\end{table}

Regarding to the downstream impacts of the ranking functions, 
we measure the average number of messages per search that searchers send to results. 
On LinkedIn, sending a message is typically a way to start a recruiting process. 
Thus, this metric is particularly important. As shown in Table \ref{table_second_order}, on homepage search, the new ranking function improves 
the metric by {\it20\%} over the legacy system and on recruiter search, the new ranking function improves {\it37\%}. All of the improvements on both first-order and downstream metrics are statistically significant. The significant improvements across all metrics confirm the benefit of 
the expertise scores and the learning approach. We further conduct additional A/B tests to verify the impact of each. The first A/B test compares the legacy system and a machine-learnt model with the same set of features (i.e. without expertise feature). The second test compares machine-learnt models with and without the expertise feature. The experimental results show both of the improvements are statistically significant.

Between homepage search and recruiter search, interestingly, the improvements achieved on recruiter search are higher than the ones on homepage search on every metric. This is probably because the users on recruiter search are more active thus they take advantage of the quality improvement more.

\section{Conclusions}
%

In this paper, we introduce the problem of personalized expertise search and point out 
technical challenges faced when building such a system at scale.
We propose a scalable way to derive expertise scores over a large corpus. 
Our techniques take into account skill co-occurrence patterns to estimate on both explicit and inferred skills. We leverage state of the art LTR algorithms to learn final ranking functions combining expertise scores with other 
non-personalized and personalized features. 
To extract training data, we introduce a deterministic top-{\it N} randomization strategy coupled with {\it easy negatives} alleviating the biases in search logs. 

We conducted A/B tests on both LinkedIn homepage search and recruiter search (a premium product) for a reasonably long period of time. The experimental results show substantial improvements over the legacy system. Specifically, on homepage search, the proposed approach increases {\bf18\%} on CTR@1, {\bf11\%} on CTR@10, {\bf14\%} on MRR and {\bf20\%} on downstream messages sent from search. On recruiter search, the approach achieves even bigger improvements of {\bf31\%} on CTR@1, {\bf18\%} on CTR@10, {\bf22\%} on MRR and {\bf37\%} on downstream messages. The machine-learnt models are serving nearly {\bf all live traffic} for skills search on LinkedIn.

We summarize some of our key findings below:
\begin{itemize}
  \item For finding experts in large professional networks like LinkedIn, it is crucial to go beyond text matching. Our approach to inferring skill expertise scores yields a significant lift. In terms of normalized feature weights, the expertise scores is amongst the top four most important features in the final ranking function (due to SEO-related issues, we cannot release the actual feature values or their order of importance).
  \item Collaborative filtering, which has been widely used for recommendation, turns out to also work well for inferring skill expertise scores in professional networks.
  \item Personalization plays a crucial role in expertise search. For instance, the best geo feature (in terms of normalized feature weights) is also amongst the top four features.
  \item Re-ranking offline performance on the training data extracted by the deterministic top-{\it N} randomization strategy coupled with {\it easy negatives} is directionally inline with online performance. 
  Between two ranking functions, the one with higher offline re-ranking performance also achieves better results on online A/B test where they are used to rank the whole sets of results. 
  Thus, this strategy is very useful for personalized training data collection and for offline testing to make decision on which models to deploy.
  \item In terms of system architecture, the two-phase approach (offline and online) is a reasonable choice for production ranking systems. 
  The offline phase runs on distributed computing platforms like Hadoop. It
  allows processing very large datasets with complex methods e.g. matrix factorization to produce sophisticated signals like expertise scores. 
  It periodically generates new versions of the signals offline. The online phase then simply consumes the latest versions of the signals in combination with other signals 
  available in the index to generate final rankings in real time.
    
\end{itemize}



\bibliographystyle{IEEEtran}
\bibliography{IEEEexample}

%
%
%

\end{document}